# A Conversation with Seymour Geisser

**Ronald Christensen and Wesley Johnson**

*Abstract.* Seymour Geisser received his bachelor's degree in Mathematics from the City College of New York in 1950, and his M.A. and Ph.D. degrees in Mathematical Statistics at the University of North Carolina in 1952 and 1955, respectively. He then held positions at the National Bureau of Standards and the National Institute of Mental Health until 1961. From 1961 until 1965, he was Chief of the Biometry Section at the National Institute of Arthritis and Metabolic Diseases, and also held the position of Professorial Lecturer at the George Washington University from 1960 to 1965. From 1965 to 1970, he was the founding Chair of the Department of Statistics at the State University of New York, Buffalo, and in 1971, he became the founding Director of the School of Statistics at the University of Minnesota, remaining in that position until 2001. He held visiting professorships at Iowa State University, 1960; University of Wisconsin, 1964; University of Tel-Aviv (Israel), 1971; University of Waterloo (Canada), 1972; Stanford University, 1976, 1977, 1988; Carnegie Mellon University, 1976; University of the Orange Free State (South Africa), 1978, 1993; Harvard University, 1981; University of Chicago, 1985; University of Warwick (England), 1986; University of Modena (Italy), 1996; and National Chiao Tung University (Taiwan), 1998. He was the Lady Davis Visiting Professor, Hebrew University of Jerusalem, 1991, 1994, 1999, and the Schor Scholar, Merck Research Laboratories, 2002-2003. He was a Fellow of the Institute of Mathematical Statistics and the American Statistical Association.

Seymour is listed in World Men of Science, American Men and Women of Science and Who's Who in America. He served on numerous committees for the National Institutes of Health, Food and Drug Administration, National Institute of Statistical Science and National Research Council. In addition, he was a National Science Foundation Lecturer in Statistics from 1966–1969; member of the National Research Council Committee on National Statistics from 1984–1987; Chair of the National Academy of Sciences panel on Occupational Safety and Health Statistics from 1986–1987; and he served on Program Review Committees for many universities. He delivered the American Statistical Association President's Invited Address in 1991.

Seymour authored or coauthored 176 scientific articles, discussions, book reviews and books over his career. One of his articles, "On methods in the analysis of profile data," which was coauthored by S. W. Greenhouse, and published in *Psychometrika* in 1959, is listed as a citation classic. He pioneered several important areas of statistical endeavor. He and Mervyn Stone simultaneously and independently invented the now popular statistical method called "cross-validation," which is used for validating statistical models. Dr. Geisser's paper on the subject, "The predictive sample reuse method with applications," was published in the *Journal of the American Statistical Association* in 1975. He pioneered the areas of Bayesian multivariate analysis and discrimination, Bayesian diagnostics for statistical prediction and estimation models, Bayesian interim analysis, testing for Hardy–Weinberg equilibrium using forensic DNA data, and the optimal administration of multiple diagnostic screening tests.

*Ronald Christensen is Professor, Department of Mathematics and Statistics, University of New Mexico, Albuquerque, New Mexico 57131, USA e-mail: fletcher@stat.unm.edu. Wesley Johnson is Professor, Department of Statistics, University of California, Irvine, California 92697, USA e-mail: wjohnson@uci.edu.*

Seymour was primarily noted for his sustained focus on prediction in Statistics. This began with his work on Bayesian classification. He gave an early exposition in his article, "The inferential use of predictive distributions," published in *Foundations of Statistical Inference* in 1971. Most of his work in this area is summarized in his monograph, *Predictive Inference: An Introduction* published by Chapman and Hall in 1993. The essence of his argument was that Statistics should focus on observable quantities rather than on unobservable parameters that often do not exist and have been incorporated largely for convenience. He argued that the success of a statistical model should be measured by the quality of the predictions made from it. He pointed out that interest in model parameters often seemed to be based more on interest in ease of mathematical display than in their scientific utility.

Shortly after it was introduced, he gave his attention to forensic DNA profiling. He was involved as an expert witness in 100 litigations involving murder, rape, paternity, and other issues. His experiences in dealing with the FBI throughout these litigations are catalogued in his paper, "Statistics, litigation and conduct unbecoming," published in *Statistical Science in the Courtroom* in 2000. His primary purpose in these litigations was to point out that statistical calculations displayed in court should be valid. It was his contention that the statistical methods then being used by the prosecution in DNA cases were flawed.

Finally, he was proud of his role in the development of the University of Minnesota School of Statistics and its graduate program. During his tenure there, he was responsible for hiring outstanding faculty who have since become leaders in their areas of expertise. Moreover, many of the students obtaining their Ph.D. degrees at the University of Minnesota have also become prominent in their respective fields. Seymour was substantially responsible for creating an educational environment that valued the foundations of Statistics beyond mere technical expertise.

Two special conferences were convened to honor the contributions of Seymour to the field of Statistics. The first was organized by Jack Lee and held at the National Chiao Tung University of Taiwan in December of 1995. The second was organized by Glen Meeden and held at the University of Minnesota in May of 2002. In conjunction with the former conference, a special volume entitled, *Modeling and Prediction: Honoring Seymour Geisser,* was published in 1996.

Most recently, Seymour compiled his lecture notes into a manuscript entitled, *Modes of Parametric Statistical Inference* [published by Wiley in 2006]. The book provides a broad view of the foundations of Statistics and invites discussion of the relative merits of different modes of statistical inference, method, and thought. His life's work exemplifies the presentation of thoughtful, principled, reasoned, and coherent statistical methods to be used in the search for scientific truth.

Seymour Geisser died March 11, 2004.

The Department of Statistics at the University of Minnesota has established the **Seymour Geisser Lectureship in Statistics**. Each year, starting in the Fall of 2005, an individual will be named the Seymour Geisser Lecturer for that year and will be invited to give a special lecture. Individuals will be selected on the basis of excellence in statistical endeavors and their corresponding contributions to science, both statistical and otherwise. For more information or to see his curriculum vitae, visit the University of Minnesota, Department of Statistics web page, www.stat.umn.edu.

This conversation took place among Seymour Geisser, Wesley Johnson and Ronald Christensen in Seymour's home in St. Paul, Minnesota on January 15 and 17, 2004. He had earlier been diagnosed with two very rare and incurable diseases. Despite his illness and discomfort, he was in good humor throughout the interview. His wife, Anne, was an inspiration to him and as he became more ill, she played the major role in his emotional and physical support. This conversation could not have taken place without her and we dedicate it to Anne.

**Wes:** Tell us something about your early life.

**Seymour:** Well, my parents were immigrants from Poland. They came here in the early 1920s—of course very, very poor. My father ended up being a garment worker in New York City. I was born in the Bronx, but have no recollection of it because at the age of two, we moved to Brooklyn. I have one brother who is four years younger than I am.

I was a student at Lafayette High School in Brooklyn and my undergraduate college was the City College of New York, which was entirely free at the time. It was quite a chore to get up to City College from where I lived. City College was up on Convent Avenue and 137th Street and I lived down in Bensonhurst, which was almost on the southern end of Brooklyn. It took almost two hours going and two hours coming back. I spent a lot of time sleeping on the subway.

**Ron:** Do you want to say any more about your parents?

**Seymour:** They came from Warsaw, Poland. My father was drafted into the Polish Army during the Russo/Polish war of 1920 and as soon as he was discharged, he left the country. My mother followed soon after. My father was 19 and I think my mother was 18 and they had just gotten married. They didn't have enough money for both to come at the same time. So, he came first and worked here for almost a year in a number of jobs. One of them was a singing waiter.

**Ron:** I'm curious why your father left right after the war.

**Seymour:** Well, they were Polish Jews and things weren't very good for them there. They were always on the brink of starvation and then there was the war with Russia. He had three brothers who, discharged from the Russian army in 1905 during the Russo/Japanese War, also left and came to this country. So, he came over and brought his mother,

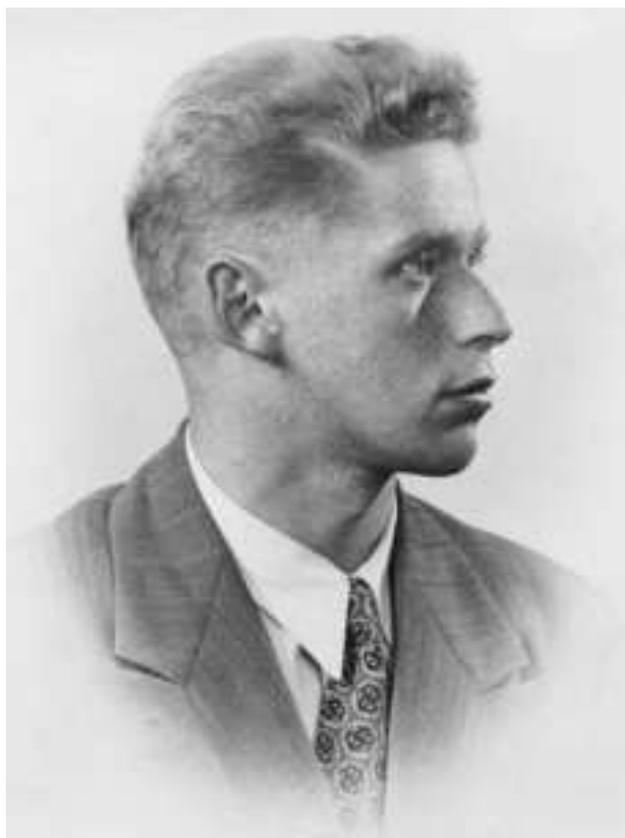

Fig. 1. *Seymour circa 1950, with red hair.*

his wife and, I believe, his sister, although I'm not sure about that one.

**Ron:** How did your father get separated from his brothers? They were in Russia?

**Seymour:** Things were very difficult in Poland. At the time it was under occupation by Russia. It was called Russian Poland. It was very difficult to work and make a living and there was opportunity in this country, or at least they thought there was [Seymour smiles], and they managed to come. One of his brothers had to stop in Belgium because he didn't have enough money to go any further. He worked as a coal miner in Belgium for a year or so and then he came over -just the usual immigration pattern.

**Ron:** Do you want to say anything about your mother?

**Seymour:** Yes. My mother had a brother who lived in the States. He and my father actually came over at the same time. I think they were both discharged from the army at the same time. My mother didn't have much schooling. My father didn't have much regular schooling, but he picked up a lot of languages. He could speak fairly fluently in, of course, Polish, as well as Russian, Ukrainian, Hebrew, Yiddish, and Spanish. Spanish he picked up from Cubans when he was working in this country. My mother was a housewife for a number of years and later, when I was in my teens, she went to work, too, as a garment worker.

**Wes:** So they all went to New York?

**Seymour:** Well yes, and then one of them, with his family, moved to California and eventually settled in the L.A. area. I have an aunt, who was my father's sister, who was the mother of Leon Gilford, who had been a statistician at the Census Bureau, and his wife was Dorothy Gilford. They are the ones that got me interested in Statistics. Dorothy Gilford had been a student of Hotelling's at Columbia, I believe, and my cousin Leon had been a student of hers, I guess also at Columbia. I was an undergraduate math major at City College. One of the reasons was [Seymour laughs] because we used to play chess in the cafeteria which was near the Math rooms and offices, so we didn't have far to go. And then they got me interested in Statistics. I really didn't know about graduate school at the time. I was sort of a novice at all this and they suggested that I apply to North Carolina.

**Wes:** How did they get you interested in Statistics?

**Seymour:** Opportunities!!! [Laughs all around.] The only thing you could do in math back then was to teach high school and I didn't want to do that.

**Ron:** Did you like high school?

**Seymour:** Oh, high school was a good bit of fun. I played a little basketball on the high school team and then I had an argument with the coach and I quit. [My parents] very much hoped I would go to a university and, of course, at the time, it was free.

**Wes:** Tell us about your graduate education.

**Seymour:** When I left City College to go to Chapel Hill, I thought I was entering a country club. It was such a pretty campus, even just the walk over to Phillips Hall where the Statistics Office was. Life seemed very easy. They had a lot of students there, many of whom have made a mark in the Statistics field. There were a few upper classmen such as Sudish Ghurye, Ingram Olkin, Ralph Bradley, Milton Terry, Sutton Munroe and others. Those closer to my time, which was two or three years behind, were people like Don Burkholder, Ted Colton, Fred Descloux, Ed Gehan, Ram Gnanadesikan, Shanti Gupta, Jack Hall, Bill Howe, Marvin Kastenbaum, T. V. Narayana, Jim Pacheres, K. V. Ramachandran, Paul Somerville, Bill Thompson, John Wilkinson and Marvin Zelen. There were others, too, but I can't recollect all their names. I think I've given you a pretty long list.

**Wes:** Was Marvin your year?

**Seymour:** He was a year ahead of me. He just took a Master's Degree and then he left for his job at the Stevens Institute of Technology.

**Wes:** Did the students go out together?

**Seymour:** Some of us did. We would do the usual things: drank beer in the rathskeller, played a lot of cards, gambled a lot. I think I held my own.

**Wes:** What were the professors like?

**Seymour:** Well, when I came there, they had an excellent entourage of professors. Harold Hotelling, Wassily Hoeffding, S. N. Roy, George Nicholson, R. C. Bose and Herb Robbins were some of the top people at the time. Robbins was a very good teacher. We didn't have too much interaction with our professors; they were all quite a bit older and the only interaction with them was when they were mentoring us as students. Hotelling was my mentor, but he was very hard to get, and every time I would find him and show him my work, he would always suggest something more to do. I got to be a little annoyed at this. I thought I had done enough. So the next time he asked me to do something, I went back and I did it and I thought, what would he ask next. I thought about it and I said probably this kind of thing and I did it. Next time I came in, sure enough, he asked me to do exactly that and I said, "Here, I've done it." He said, "Well then, I guess you're finished."

**Wes:** Why did Hotelling stand out for you?

**Seymour:** Hotelling had done all this work in multivariate analysis, T-square, canonical correlations, discrimination. He was very good in small sample theory. As I say, my cousin's wife had worked under him and she said he was the top man. So, I took my Master's thesis with him and my Doctoral dissertation.

**Ron:** Just briefly, what were they?

**Seymour:** The Master's thesis was on matrix computations: roots and characteristic vectors of matrices. The Ph.D. thesis was on the mean square successive difference in Statistics. During the summers of 1952 and 1953, when I was a graduate student, I worked at the Aberdeen Proving Ground. John von Neumann had worked for them previously during the war, wrote some papers with some of the people there on certain quadratic forms. I got interested and did some work on that, which is really two parts, mean square successive difference as an estimate of the variance when you have a slowly moving trend—so you can difference it out. And the second part is the ratio of that to the sample variance, which was the statistic that von Neumann worked on, which apparently was useful on a firing range. So I picked that problem and Hotelling said, "Fine, work on it."

**Ron:** I think successive differences are used in control charts, aren't they?

**Seymour:** Yes, when you have things that are related and they have a linear trend. There used to be a guy at Iowa State who did a lot of work on that sort of thing. [See Rao (1959).]

**Ron:** I assume that when you say "Aberdeen," you don't mean Aberdeen, South Dakota.

**Seymour:** No, it is Aberdeen, Maryland. It was a proving ground where they tested ammunition, guns, but they had a cannon mounted on a platform between two trucks. We also had a lot of recoilless rifles that were tested. That was 50 years ago. Sometimes, when a deer used to cross the path of the target they used to turn the cannon on the deer.

**Ron:** Did they ever hit the deer?

**Seymour:** I don't think so.

**Wes:** Why did North Carolina attract so many famous people?

**Seymour:** Because of Hotelling. He was brought down by Gertrude Cox from Columbia where they didn't have a Statistics Department at the time. He was actually in Economics, I believe. In fact, he had done very important work in Economic theory before he came to North Carolina. He attracted the two people from India, Bose and Roy, and Hoeffding. I think Hoeffding was in Germany during the war. I'm not sure whether he was a Dane originally and a refugee, but he came over. [Ed. Note: Hoeffding was born in Finland, moved to Denmark in 1920, and to Germany in 1924.] And then, of course, Herb Robbins. Robbins left in about 1952 or 1953 to go back to Columbia.

Hotelling and [Jerzy] Neyman were the key figures in Statistics at that time, the late 1940s to mid-1950s. [Hotelling] actually put out the idea of a confidence coefficient, a confidence interval, in a paper which predated Neyman's work. But he never developed it as Neyman and [Eagon] Pearson did.

**Ron:** Did any interesting people come by to visit North Carolina?

**Seymour:** C. R. Rao came by, when he was quite young. We had some economists, Wassily Leontiff and others. Wally [Walter L.] Smith. Somebody came who taught a course I took in Actuarial Statistics. But it wasn't very interesting.

R. A. Fisher spent the summer, I think, at North Carolina State. There was an interesting story about Fisher. He was at a picnic, if I remember this correctly, celebrating the 4th of July. Some woman came up to him and asked him if they also celebrated the 4th of July in England. He thought for a while and said, "Maybe they should."

**Wes:** You went to work at NIH after you got your degree?

**Seymour:** Actually, I went to work first at the National Bureau of Standards and I was there for about six months. I really wasn't interested in teaching, although I had an offer from [the Math department at] Illinois. They had some statistics vacancies. The pay wasn't very good—$5,000 or something. I went to the Bureau of Standards because it was in Washington D.C. I had lived in Washington for a summer after graduating City College. I worked at the Operations Research Office which had a branch in the Army War College. I worked there as a sort of an assistant to Herman Chernoff who was there also at that time. The Washington suburbs seemed like a nice place to live and the pay was pretty good in the civil service. There were some interesting people there. They had Jack Youden, Churchill Eisenhart, Marvin Zelen. Bill Connor was there and Bill Clatworthy.

When I was at the Statistical Engineering Lab, which Churchill Eisenhart headed, [my work was] mostly consulting and cleaning up my dissertation for publication and some other things like that. I was only there for six months. Norman Severo, I don't know if you remember him, he came two months before I left. It was pretty good; pretty interesting.

Then I heard about the Commissioned Officers Corps of the US Public Health Service and I applied and was accepted, and I was assigned to NIH [National Institutes of Health]. Actually, there was an agreement. If I would get commissioned, they would take me on. At that time, they had a uniformed service that gave you naval ranks and you were commissioned. You only wore a uniform if you were in the Public Health Service that ran the quarantines. They ran the Medical Corps for the Coast Guard and they ran the CDC, the disease center for NIH. I was commissioned as the equivalent as a Lieutenant (j.g.). I rose in the ranks to Lieutenant Senior Grade.

[NIH] was in Bethesda, Maryland. I was put in Sam Greenhouse's section in the National Institute

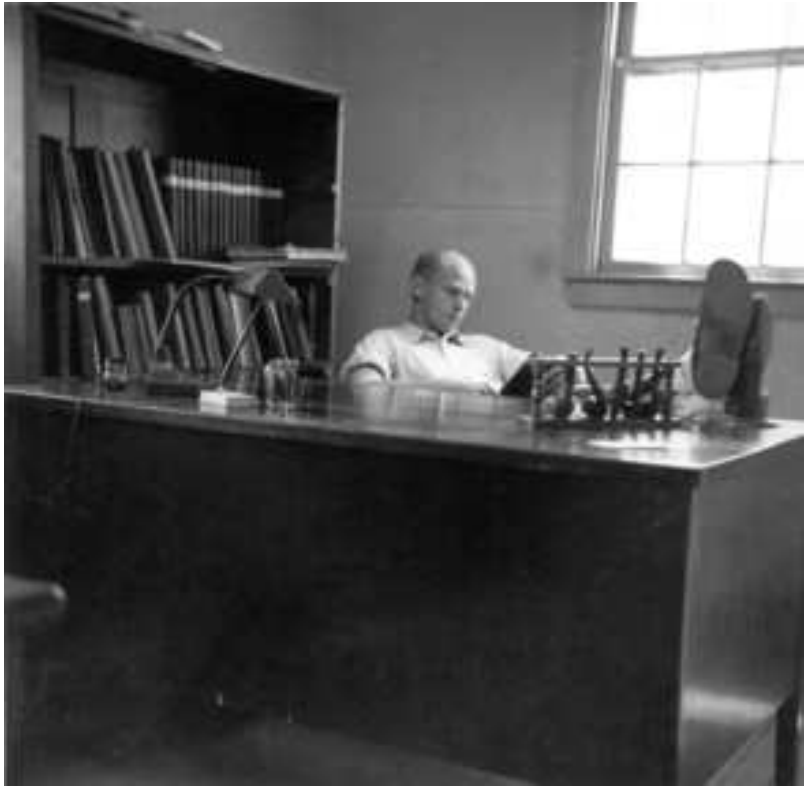

Fig. 2. *Seymour circa 1960.*

of Mental Health. [Ed. Note: Sam had agreed to support Seymour's application.] At that time Jerry Cornfield, Max Halperin, Nate Mantel, and Marvin Schneiderman were at NIH. So, there was a coterie of statisticians. Some of them were centrally located under Harold Dorn and others were at various institutes.

**Wes:** That's quite a list.

**Seymour:** Actually, it was a very interesting time there. We used to eat lunch together and talk about everything from history to statistics to religion to politics. The table talk was really interesting. Jerry Cornfield was a bit of a raconteur. But they're all gone. He was a very interesting man, Jerry. He had a Bachelor's Degree in History and the rest he learned on his own. He worked for the Bureau of Labor Statistics for a while and then he was brought over to NIH. Some of the table talk was about Bayesian credible intervals.

**Wes:** Was there any of that in North Carolina?

**Seymour:** Bayes? Long gone. The only mention was what was in [Harald] Cramér's book about "The Man in the Iron Mask." The probability of whether he was the king of France. Hoeffding was in nonparametric inference. Both Bose and Roy, at the time, more Roy, were in multivariate analysis, and Bose had just started to do work in coding theory. But there was nothing about Bayesian theory at the time. That was a subject they all thought had died in the last century.

When I told Hotelling several years later, while I was at Buffalo, that I was doing Bayesian work, he said, "Well, that comes and goes." It's interesting that Hotelling, who was a man of immense erudition and knew all sorts of things and was also a wonderful raconteur, never thought about Bayesian theory— only to make that remark to me once, that it comes and goes.

**Wes:** Was Jerry Cornfield a Bayesian when you first met him or was he becoming one?

**Seymour:** I think he was sort of leaning that way. He was very interested in exploring the differences between Bayesian and Frequentist theory. That happened while I was there. I think it was from reading [L. J.] Savage. Jerry was thinking about the fundamental ideas of gambling, like [Robert] Buehler was doing here [at Minnesota] at one time. The meaning of confidence intervals, and the meaning of what he called, at that time, credibile intervals.

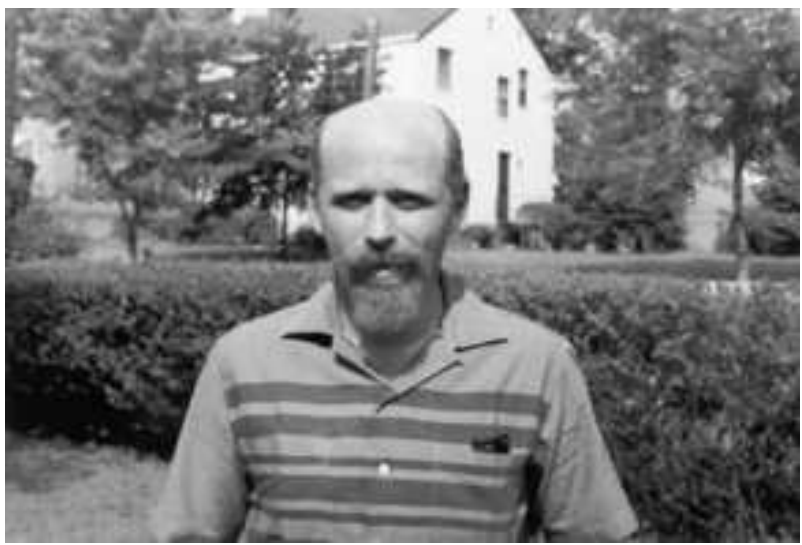

Fig. 3. *Seymour circa 1965, with red beard.*

**Wes:** You said you were talking with this larger group?
**Seymour:** None of them was really interested. Max Halperin was not interested in Bayesian theory; he pooh, poohed it. And Sam Greenhouse wasn't interested in Bayesian theory and, of course, Nate Mantel was not. They were interested in methodology mainly, in developing methodology for experiments and trials that were run at NIH, and were using the usual frequentist theory to analyze them. The NIH group was really a very smart, very good, very knowledgeable group, even though their actual statistical training was pretty minimal at that time. Only Max Halperin had a doctorate and that was from Chapel Hill. The arguments there were passionate.
**Wes:** So you leaned one way and they leaned the other?
**Seymour:** I was leaning towards the Bayesian approach, especially when I did a paper. First, I wanted to see what would happen if I considered the usual multivariate problems from the Bayesian point of view in about 1963. And then I turned to see what would happen with classification/discrimination. Then it dawned on me that with Bayesian theory you didn't have to make a [strict] separation for linear discrimination. For example, everything on one side was guilty and the other side innocent, if you like. [A Bayesian] could find the probability of each individual being one or the other. It was a much finer distinction than using, say, the usual Fisher linear discriminant and that really swung me to the Bayesian approach at the time.
**Wes:** That sounds like the beginning of thinking about prediction.
**Seymour:** Yes, of course; it is prediction or a retrodiction to whatever you want to call it. David Cox refused to let me use [retrodiction] in a paper. By a retrodiction, I mean something that has already happened but you don't know what it was, so you have to have a probability that it happened. Prediction is something that is to happen in the future.
**Wes:** What kind of problems did you work on at NIH?
**Seymour:** Well, at NIH, with Sam Greenhouse, I wrote my most infamous paper. "Infamous," I say because it wasn't a very important or very hard paper. It was just a paper that seemed to have caught on with social scientists and some medical people. It was just this profile analysis paper which ended up being a citation classic, which means that it had a lot of citations. It still has more citations than all of my [other] papers altogether. [Laughs all around.]
**Wes:** This was the Greenhouse–Geisser paper?
**Seymour:** There are two papers. The first paper was in the *Annals*, which actually worked out all of the quadratic forms, their expectations, and the mathematics. And the second was in *Psychometrika* and that was the citation classic. That was just to show the methodology—how to use this. It wasn't a very big deal. I worked much harder on other papers and I think I produced much better work. But *de* *gustibus non disputandum est* [there's no accounting for taste].

**Wes:** We skipped over your interest in Latin when you were younger.

**Seymour:** I took Latin in high school for two years and then I continued in college for another two years. I also took German and French [but] Latin is quotable.

I [also] had read a lot of history. It was one of my side interests, reading about history and archeology, and of course, novels at the time. When I was working at NIH, I would read at least a half a dozen books a week. When I was in Chapel Hill, I got interested in the Civil War. I read a lot about that. In fact, in their library they had the home of the US government published volumes on what they called the War of the Rebellion. I'd go through them; there must have been at least 20 volumes. That was the interesting thing about the Civil War; it was called the War of Rebellion in the North and, as you moved South, it was called the War for Southern Independence.

[Seymour and Ron digress about history.]

**Wes:** We've got a lot of questions, so we need to move on. You were also associated with George Washington University.

**Seymour:** Yes, George Washington University ran a Statistics Graduate Program at night. It was one of the earliest ones; it was started in the 1930s. At that time, Sam [Greenhouse] taught there, I taught there, [Solomon] Kullback taught there. Kullback was an interesting man, too. He was sort of unappreciated in his time. I remember that many, many years later, there was a meeting in his honor and I decided to give a talk on the Kullback/Leibler divergence and all its varied applications. At the time that I was teaching [at GW], I had no interest in that. At any rate, after my talk, Kullback came over to me and said, "Seymour, where the hell were you when I needed you."

**Wes:** Somewhere in there you must have met and married your wife and had children and moved across the street from Sam Greenhouse.

**Seymour:** Right. I was married in Chapel Hill to my first wife. We were married for about 22 1/2 years. We were divorced in Minneapolis, Minnesota. About four or five years later, when I was visiting the Harvard School of Public Health for a couple of months, I met Anne and we got married the following year. [Ed. note: Marvin Zelen had asked Anne, who was his administrator at the School of Public Health, to make sure Seymour was looked after.]

[My] children were all with my first wife. One was born in Washington, D.C. and three in Maryland. My oldest was a daughter and she was born in 1957. Then, I had a son in 1958, a daughter in 1960, and a son in 1963.

**Wes:** So you decided to move to Buffalo? You were attracted to snow?

**Seymour:** [Chuckle] No, it's interesting; it was, in a sense, a little bit of a mistake. I was running a section at the time in the National Institute of Arthritis and Metabolical Diseases. They were trying to reorganize their institute. Bill Clatworthy, who was the Chair of a Statistics group in the math department, tried to recruit me and made me a rather good monetary offer. During the negotiations he resigned as the Chair. I made it a condition that it be made a separate department of Statistics and they appointed me the Chair, and I went there. I also had the thought, which was not true, that since things were so interesting in the table talk around NIH, if I went to a university and had a much wider group of people to talk to, it would be even more interesting. That turned out to be completely false. It depends on the individuals, not on the place you are at necessarily.

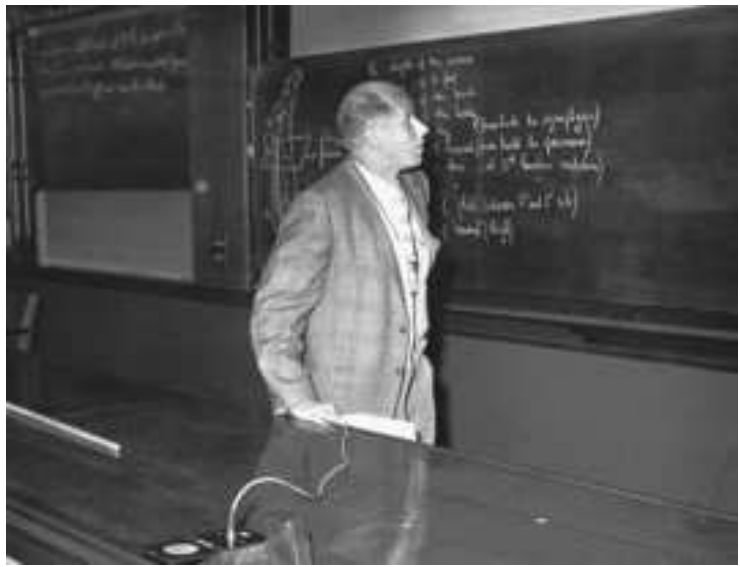

Fig. 4. *Seymour circa 1970.*

**Ron:** Who were the statisticians at Buffalo when you were there?

**Seymour:** When I was there, Norman Severo was there and he also was instrumental in bringing me. Bill Clatworthy. I think they were the only two at the time. And then when I came we hired a bunch more. There were quite a number of statisticians passing through there at one time or another, including Marvin Zelen, Manny Parzen and Charles Mode. Jack Kalbfleisch was there for a while and we had good relations with Waterloo: Dave Sprott, the Kalbfleisch brothers, and Ross Prentice who was the last person I hired just before I left. And Peter Enis was there. [Ed. Note: Peter was Seymour's first Ph.D. student at George Washington.] We had very good relations with the math department, which is very unusual. In fact, at one time, I was asked to be the Chair of the math department as well, because they had some big fights. But I told the Dean where to go.

[Living in Buffalo] wasn't too bad. We bought a house in Williamsville, which was a suburb. Eventually, the department moved from inside the city to Amherst, which is another suburb. It was easy to get over there and the parking was good. Things were easy.

**Wes:** Where is prediction at this stage in your thinking?

**Seymour:** I left NIH in 1965. It was well into my NIH days, the late 1950s, early 1960s, when I worked on the classification problem, which was essentially a prediction problem. And then finally, in 1970 at a conference at Waterloo, there I really wrote about what I thought was prediction. I wrote, "The inferential use of predictive distributions," which is in the volume, *Foundations of Statistical Inference*, edited by Godambe and Sprott (Geisser, 1971).

**Wes:** That was a lively paper. It was the first paper of yours I read.

**Seymour:** Yes, Jack [I. J.] Good commented on it. He had said that he had heard of a paper by [Bruno] de Finetti also arguing about prediction and observablistic inference rather than parametric inference. Jack commented on it, Dave Sprott, and a couple of others.

Interest seemed to be passive. People were interested in parametric inference, period.

**Wes:** What did Cornfield think about prediction?

**Seymour:** It's interesting. It wasn't discussed much there. He was still concerned about the whole Bayesian idea. He got more interested in some of the sample reuse work that I did later on, sometime in the mid 1970s. Two years before he died, he said he was thinking about using it because he was trying to make predictions about certain health problems, [for which sample reuse] seemed to him to be a good idea.

**Ron:** Your interest in Bayesian ideas seems to have developed very quickly. You got interested in it from discussions. You did the multivariate paper, then you did the classification paper and, pretty much from there, all the prediction ideas came.

**Seymour:** Yes, yes. I wasn't thinking about prediction before I did the Bayesian work. It got me thinking about the prediction of observables rather than the parameters that you never see... I talked a little bit about [frequentist prediction] in the 1971 paper, "Inferential Uses of...," in the volume edited by [Godambe] and Sprott. I thought that was a dead end because it didn't give you probabilities. Usually, when I teach it, I show how you can actually do some of the frequentist methods, such as how you can get a prediction interval for future observations, but they are so limited and so circumscribed compared to what you can do as a Bayesian, it is uninteresting.

**Wes:** You enjoyed being an administrator at NIH and Buffalo?

**Seymour:** Well, I was a petty functionary at NIH, you might say, but at Buffalo, I did run a department. In fact, through my whole university career, I was an administrator until I stepped down from the directorship at the School of Statistics at Minnesota.

**Wes:** What were your thoughts about building a group in Buffalo and then Minnesota?

**Seymour:** Building a group is sort of difficult. There are lots of ups and downs. When you have money, so does everybody else. So you are competing for some very good people. That becomes a difficult chore. And when you don't have money, you can't hire anybody. So a lot of time is spent haggling and fighting with deans about lines, space, money: the usual trinity.

**Wes:** One thing that is really noticeable to me is that you focused on hiring some amazing people.

**Seymour:** Yeah, I think we had a rather good group at Minnesota. Early on, I hired Steven Fienberg, Kit [Christopher] Bingham, David Hinkley, Don Berry, David Lane.

**Wes:** Dennis Cook, Joe [Morris L.] Eaton.

**Seymour:** Right. And Dennis has really been a fireball in the areas he is interested in. And then there were the senior people that we hired, like [Glen] Meeden and [Doug] Hawkins, and later on a rather good hire was Charley Geyer whose computational skills are really excellent. Little by little, some of the better ones got better offers elsewhere, which we ended up not being able to match or exceed and they left, such as Fienberg and Hinkley. When I was there, of course, Bob Buehler was there and Milton Sobel was there, Bernie Lindgren, Somesh Das Gupta and [Michael] Perlman, so it was a fairly strong group right up to more or less now.

**Ron:** I remember we had a lot of interesting people come to Minnesota. I remember Cochran, Rao and George Barnard.

**Wes:** Dennis Lindley came several times.

**Seymour:** There was the Physicist, Ed Jaynes.

**Wes:** I think you had a whole quarter on Fisher's contributions.

**Seymour:** That's right and we wrote a little booklet on Fisher's contributions. [Ed. note: *R. A. Fisher: An Appreciation* (Fienberg and Hinkley, 1980).] It was a very interesting time. As far as Statistics goes, it was one of the most interesting times that I had. But, on the other hand, there was sort of this lack of people going to lunch together. I ended up usually lunching with some of the philosophy profs, history, biology, and so on. I would often eat with them because most of the statisticians didn't go out to eat at the faculty club.

**Wes:** We got slightly ahead of ourselves in your transition to Minnesota. How did that happen? They made you an offer you couldn't refuse?

**Seymour:** Actually, I refused it the first time. Then they had somebody else who they made the offer to. He refused it and then they came back to me again the second time. The second time I decided to take it.

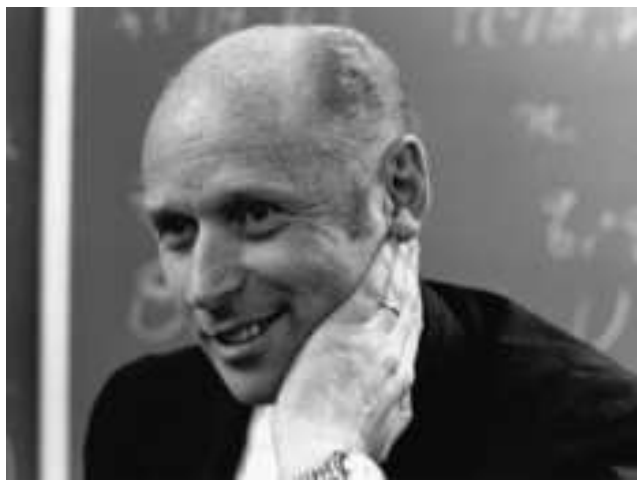

FIG. 5. *Seymour circa 1975.*

It was a chance to work in a better department at a university that was better than Buffalo was. And change. [Ed. note: Marvin Zelen has pointed out that Seymour intensely disliked the administration at Buffalo.] I had finished my term as Chair at Buffalo and we brought Manny Parzen in as Chair and I had gone on administrative leave. I was at the University of Tel Aviv. I had some telegrams back and forth and finally I decided not to go back [to Buffalo].

**Wes:** You spent most of your career at Minnesota?

**Seymour:** Yes, well, there was a time when it was very, very interesting. There was a group who was interested in the foundations of Statistics, such as Bob Buehler, David Lane, Bill Sudderth, myself, and Jim Dickey. We brought in of a lot of people who were interested in foundations. We had a lot of seminars on it and there was a lot of interesting work that was done on foundations at that time. That was, in a sense, more interesting than methodology.

**Ron:** From a Bayesian point of view, methodology was more limited back then because computing wasn't as sophisticated.

**Seymour:** Yes, people don't seem to be much interested in foundations. As long as they can compute something, they do it. [Foundations] leaven the subject, it makes it more interesting, otherwise, it's just a trite engineering problem.

**Ron:** I tend to think the most important ideas in Statistics are not terribly mathematical.

**Seymour:** That's certainly true. There are three parts to this Statistics enterprise. The mathematics that the Berkeley people really did with zest—the inferential things, and foundations, and computing. And, of course, the inferential aspects and computing are tied up with methodology.

**Wes:** Who were your primary mentors and influences?

**Seymour:** Well, Harold Hotelling was a mentor and George Barnard. Those were the two most influential people. And then, Jerry Cornfield.

I was particularly influenced by George Barnard. I always read his papers. He had a great way of writing. Excellent prose. And he was essentially trained in Philosophy—in Logic—at Cambridge. Of all of the people who influenced me, I would say that he was probably the most influential. He was the one that was interested in foundations.

**Wes:** What are your other interests as a Statistician?

**Seymour:** I got interested later on in what I call predictive sample reuse to do predictions without stochastic assumptions and I wrote a number of papers on that. This also mirrored the work of Mervyn Stone in London. He was working in that area, too. It was motivated by trying to validate something. I wanted to work with Kit Bingham to do that because he was our computer expert at the time, and he didn't seem to be interested, so I had to work out a method that would not necessarily involve computing. Then it turned out that Stone was doing the same thing in London.

**Ron:** Multivariate analysis, prediction, Bayesian statistics, predictive sample reuse. Anything else?

**Seymour:** Well, I got interested in some legal problems. DNA. The use of statistics in DNA. It is hard to remember these things; it has been a lifetime. Then there were medical problems that I got into, and clinical trials. Oh, model selection! And the work with Wes on influential observations. More recently, there was the work with Wes on diagnostic tests. And then I got interested in comparing different diagnostic tests and screening procedures. And more recently, I also started with Wes on the optimal administration of multiple screening tests – how you did this through decision theory. I also got interested, with a student, George Papandenatos, in Bayesian interim analysis with lifetime data. That is pretty much where I am now. My research has sort of tapered off.

**Wes:** What sparked you to look into the DNA controversy?

**Seymour:** I was called up by a public defender in the county to help him sort out the DNA evidence against his client. I was one of the few statisticians involved. Most of the people involved called themselves—statistical geneticists. So I got a lot of calls. Too many calls. So I had to avoid them.

There were several [controversies] at the time. I think they use different methodologies now. At the time, there was the match, and that was either a yes or no according to the lab. The match was not necessarily an exact match, but it had to be within certain statistical tolerance levels which they decided upon. That was the first thing. And the second thing was computation of the relative frequency of individuals in different racial categories having that particular constellation of DNA, which was at the time anywhere between three and eight different markers. That was also subject to controversy. The assumption was that all the markers were independent, so that they could just multiply the probabilities that they calculated to get the final probability. You [Wes] and I showed that it was not entirely true. There was dependence. We had some papers published in the *American Journal of Human Genetics*.

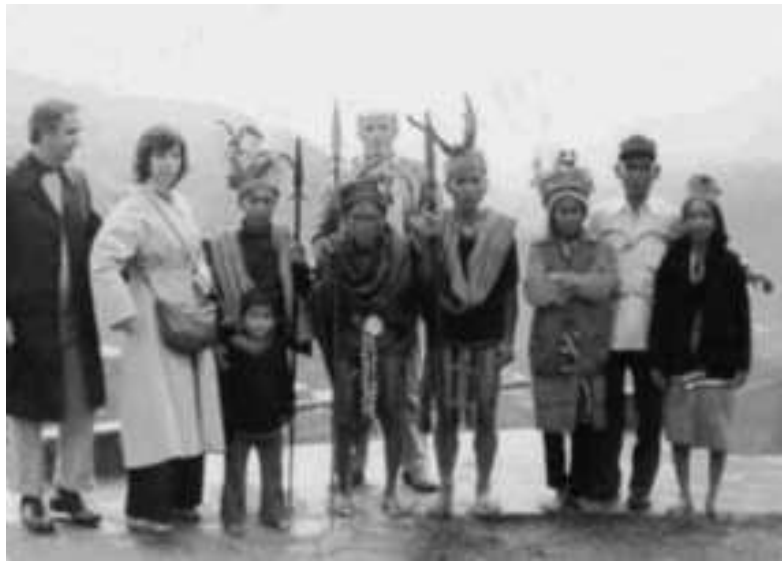

Fig. 6. *Seymour with Jim and Grace Press in the Phillipines, 1979.*

The FBI tried to stop the publication of certain papers. Not only of mine, but other people as well who decided to chime in on the problems. Then I wrote a long paper, well, not a long paper, but a paper on statistics litigation, "[Statistics, litigation, and] conduct unbecoming"—which outlined all the errors and the nastiness of the FBI.

It was published in Joe Gastwirth's volume on legal statistics. [Ed. note: *Statistical Science in the Courtroom*, Springer (Geisser, 2000).]

**Wes:** What do you see as your main accomplishments?

**Seymour:** Well, my work on Bayesian analysis and prediction. It is things that flowed from that, like model selection, diagnostics and interim analysis. I think those were my main interests and contributions. Oh, yes, administration was an interest of mine. Certainly when I was younger and had the fighting spirit to call down Deans and argue with them all the time until I wore them out or they wore me out.

**Wes:** And what do you see as the major historic trends for Statistics?

**Seymour:** If I knew what the major trends in the future would be, I'd go and do them, but unfortunately, I don't know. We seem to have a change in direction, roughly, every 15 years or so. The last change was the computer revolution, but what will come next, I have no idea. But it will come from one of the younger people, not one of us older people who continue in the same direction.

**Wes:** Please speak about the role of prediction in Statistics and more generally in science.

**Seymour:** It always seemed to me that prediction was critical to modern science. There are really two parts, especially for Statistics. There is description; that is, you are trying to describe and model some sort of process, which will never be true and essentially you introduce lots of artifacts into that sort of thing. Prediction is the one thing you can really talk about, in a sense, because what you predict will either happen or not happen and you will know exactly where you stand, whether you predicted the phenomenon or not. Of course, Statistics is the so called science of uncertainty, essentially prediction, trying to know something about what is going to happen and what has happened that you don't know about. This is true in science, too. Science changes when predictions do not come true.

**Ron:** A lot of people think that science is about understanding phenomena. In reality, what people take to be understanding, which may be correct or incorrect, helps them to develop models. The real test is whether these models help you predict correctly or not.

**Seymour:** You said it better than I did.

**Ron:** Well, I got the ideas from you.

**Seymour:** [Smiling.] Obviously, I've forgotten some of them.

**Wes:** Tell us your thoughts about Fisher's contributions to Statistics and how they contrast with Neyman's.

**Seymour:** Fisher was the master genius in Statistics and his major contributions, in some sense, were the methodologies that needed to be introduced, his thoughts about what inference is, and what the foundations of Statistics were to be. With regard to Neyman, he came out of Mathematics and his ideas were to make Statistics a real mathematical science and attempt to develop precise methods that would hold up under any mathematical set up, especially his confidence intervals and estimation theory. I believe that is what he tried to do. He also originally tried to show that Fisher's fiducial intervals were essentially confidence intervals and later decided that they were quite different. Fisher also said that they were quite different. Essentially, the thing about Neyman is that he introduced, much more widely, the idea of proving things mathematically—in developing mathematical structures into the statistical enterprise.

**Ron:** My acquaintance is limited, so I like to say that I have never met anybody who understands fiducial inference. Can you comment on that? Do I just need to meet more people?

**Seymour:** When you say you don't understand it, do you mean you don't understand what it does or you don't understand how it is defined?

**Ron:** What I am thinking of is the rationale of turning the one probability into the other.

**Seymour:** I think that turned out to be a mathematical trick, a mathematical device rather than a fundamental philosophical methodology. You have the distribution function of the statistic, given the parameter. Now, if you look at that and see that if you let the statistic be constant and vary the parameter, and if that now looks like a distribution function, Fisher essentially called that a fiducial distribution function. Fisher used this as an inversion without a Bayesian prior. It is so restricted, as is shown by [Dennis] Lindley, it is restricted to the normal-gamma family and that is the only place you can use it.

**Ron:** Bob Buehler did a lot of fiducial stuff didn't he?

**Seymour:** Oh, yes. He certainly was able to interpret Fisher correctly. Fisher's writings were always fuzzy. Buehler wrote a technical report and a set of notes that set out exactly what Fisher did in terms of the inversion. The inversion makes a big jump in that something that was a constant suddenly becomes a random variable.

**Wes:** If somebody other than Fisher had come up with this, would we still be talking about it?

**Seymour:** Probably not.

**Ron:** I do find it kind of funny that people get so wrought up about parameters being fixed constants when models are merely approximations to reality to begin with. I don't know why you'd think that the parameters you use in those models are somehow determined by a higher intelligence.

**Seymour:** There's something to be said for that. Actually, people get hung up even further by thinking the parameters are real when they're artifacts of our minds. The only time they are real, in a sense, is when you take a statistic that is based on, say, $n$ observations and then you let $n$ go to infinity. That can also define a parameter.

**Wes:** Did you meet Fisher and Neyman?

**Seymour:** I met Fisher, when I was at the National Bureau of Standards; he came there and gave a series of talks. He was one of the worst lecturers that I have ever seen. He would look only at the board and write very small and talked to the board, showing almost contempt for the audience.

With Neyman, when I was at NIH, he came to NIH for money for the Berkeley Symposium, and then went back and he said, "I asked for delta and I got epsilon." But he came to NIH and gave a few talks and one of the talks was on astronomy, as I recall. We had a guy who was actually an astronomer at NIH at the time and had changed his interests to Biophysics and he was quite interested in what Neyman was saying. At that time he was talking about work by Neyman and [Elizabeth] Scott, which involved a series of papers on the material in the heavens, distributed random or not, and tests for that. It's interesting work. He also discussed stochastic processes of diseases and the one-hit/two-hit theory of cancer.

**Wes:** What about [Harold] Jeffreys?

**Seymour:** Jeffreys had a quite different view of probability and statistics. One of the interesting things about Jeffreys is that he thought his most important contribution was significance testing, which drove [Jerry Cornfield?] crazy because, "That's going to be the least important end of statistics." But Jeffreys really brought back the Bayesian point of view. He had a view that you could have an objective type Bayesian situation where you could devise a prior that was more or less reasonable for the problem and, certainly with a large number of observations, the prior would be washed out anyway. I think that was his most important contribution— the rejuvenation of the Bayesian approach before anyone else in statistics through his book, *Probability Theory* (Jeffreys, 1961). Savage was the one that brought Bayesianism to the States and that is where it spread from.

**Wes:** Distinguish between Jeffreys' and Savage's roles in bringing Bayesianism back.

**Seymour:** Well, Jeffreys's book was always hard to read and I don't know how many people actually read it thoroughly. It probably was based on earlier work of the young economist [F. P.] Ramsey, who died young at the age of 26, I think. But, essentially, Jeffreys carried that torch alone, because all of the other statisticians in England were much more influenced by Fisher. Savage worked with de Finetti for awhile in Italy. After he worked with de Finetti he became a committed Bayesian and he preached Bayesianism. Savage was a charismatic speaker and excellent lecturer and I think that he is the one who essentially reintroduced Bayesianism in the United States and I think that is when it took off.

Very little was known about de Finetti in the States until his book was translated, the one on probability, in the early 1970s. He wasn't a well known name. [Ed. note: *Theory of Probability*, Vols. 1 and 2, Wiley. De Finetti (1974, 1975).] [His] view that it was the observables that were important, not the parameters really, caught on here with Savage. Jeffreys was probably better known, but it seems to me he was more or less disregarded in England because of Fisher's eminence.

**Ron:** It is interesting. I think you said that Ramsey was an economist. Jeffreys was a geophysicist. De Finetti was an economist, right?

**Seymour:** Actuary.

**Ron:** And Savage's book (Savage, 1954) certainly tends toward the economic, with utilities and all. Statisticians don't seem to have much input here.

**Seymour:** Well, in those days there weren't very many statisticians. Most of the statisticians were involved in government statistics. That's what statistics means—affairs of government—and were not into the sciences. I guess it was Fisher who first got into the natural sciences, that is, as opposed to the social sciences. He got into genetics and biology and agriculture, and that's what started it. Interesting. You might call it scientific statistics.

**Wes:** Can you weigh in on the Fisher/Neyman controversies?

**Seymour:** Well, at first it seems to me, they thought they were talking about the same thing and then it turned out they weren't. Fisher brooks no interference from anybody and apparently he got angry at Neyman and from then on it was a fight between the two of them on these different views of how to make inferences in statistics. Fisher claimed that, at best, Neyman's views were only good for things like quality control, not really a methodology for science. And of course Neyman disagreed. Neyman started to work on statistics related to health problems, astronomy, and in the substantive sciences. Fisher always seemed to be involved in the substantive sciences.

**Wes:** He was famous in genetics, as well as in statistics. Maybe statistics was something for him to use to do genetics.

**Ron:** I've met geneticists who didn't realize that Fisher was a statistician.

**Seymour:** Yes, that's true, but I think the interest in statistics started with Fisher. He was an excellent applied mathematician, Fisher. He may have been a student of [G. H.] Hardy's at Cambridge. His first forays into statistics, I think, were in proofs, which he saw in his mind actually, because his proofs were not really well laid out, of the distribution of the correlation coefficient, distribution of Student's *t,* so he was involved early in the distribution of small sample statistics. The Fisher/Pearson Controversy started out with the two by two table. [Karl] Pearson thought it was three degrees of freedom, and then Fisher showed pretty conclusively that it was only one degree of freedom in the analysis of the significance test for a two by two table. So he got into his first big argument with Pearson when he was quite a young man, Fisher. And then of course Pearson's son, Egon Pearson, worked with Neyman, and that was probably enough to set Fisher on fire against Neyman. He was a man of large temper. It's because, I think, he had red hair. That's true; he had red hair. Even if he didn't, you would always see him as having red hair.

**Ron:** Were you a person of large temper, because you had red hair?

**Seymour:** When I was a young and had hair. Now I'm old and very mellow.

**Wes:** Other than your own, what are some of your favorite statistics books and papers, and what makes them your favorites?

**Seymour:** Well, my two favorite books, that I look at quite frequently, are Fisher's *Statistical Method in Scientific Inference* (Fisher, 1956) and Cramér [*Mathematical Methods of Statistics*] (Cramér, 1946). Those are the two books that I've learned the most from. The one, Cramér, for the mathematics of Statistics, and from Fisher, thinking about the philosophical underpinnings of what Statistics was all about. I still read those books. There always seems to be something in there I missed the first time, the second time, the third time.

The papers I always enjoyed the most were those written by George Barnard. There was a whole sequence of papers and those are the ones I also reread occasionally.

**Wes:** What advice would you give new researchers in statistics?

**Seymour:** Forget about your dissertation, and try to strike out in different directions. Because your dissertation is normally, at least currently, almost dictated by your advisor, who told you what area to look at, where to go, and so on. If you really want to do something important in statistics, you better strike out on your own, take a chance. That's not easy in the American system where you have to get tenure. The easiest way to get tenure is to continue on the same course that you were on in graduate school and continue pumping out papers in that area. But if you want to live a little more dangerously, try to strike out on your own. That is where the new stuff, the innovations that can change the direction of statistics, come from. The younger people have many very bright and innovative ideas.

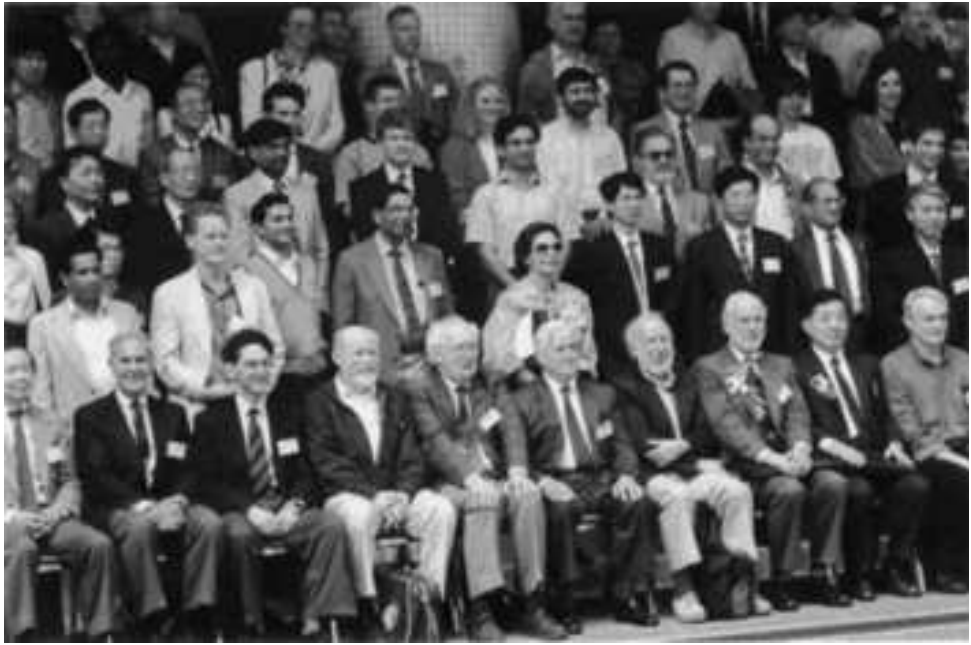

Fig. 7. *Seymour and Colleagues—Hong Kong—March 1992. Symposium on Multivariate Analysis.*

**Wes:** Please tell us your vision of the future.
**Seymour:** I really don't know. My own feeling of where it should go is in the direction of talking about and inferring about observables rather than parameters. An important area [is] developing new models for scientific work. We have a very limited number of volumes that we pull off the shelf for almost any problem that occurs. I am sure ten years from now, or even less, there will be a quite different view of where statistics is and where it should go.
**Ron:** I must have gotten those ideas from you, because I feel very strongly that both of those are what we should be doing.
**Wes:** Leaving the subject of Statistics, you once told me that you would like to be a writer. With sev-eral books and numerous papers, in a sense you are. What would you have written if not Statistics?

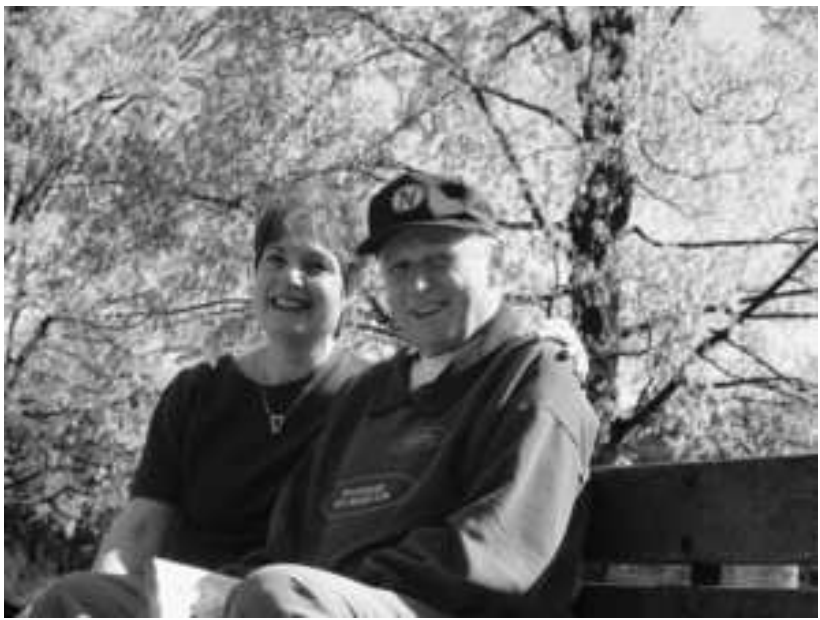

Fig. 8. *Seymour and wife Anne—Central Pennsylvania, 2003.*

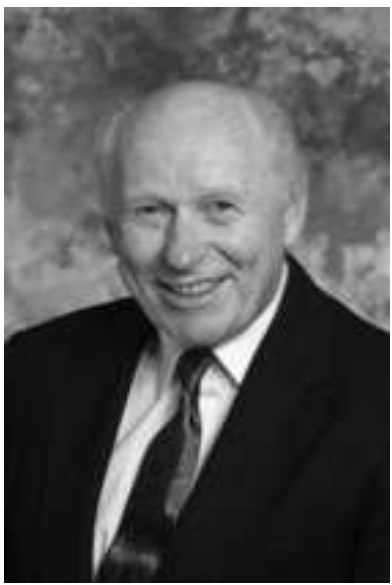

Fig. 9. *Seymour Geisser: Director, School of Statistics, University of Minnesota, 1971–2001.*

**Seymour:** Well, as I said before, I sort of backed into Statistics, because at the time I got my mathematics degree there wasn't much you could do with it. And I had these cousins who were statisticians and told me that it looked like a good enterprise and you could actually eat if you were a statistician, so I went to graduate school. But the things that I was also interested in were history, archeology, religion, novels. I was especially interested in biblical archeology. I subscribed to a biblical archeological journal and history always fascinated me, any kind of history, ancient history, medieval history, just what has happened to man. I was once interested in writing novels. But you get so tied down in your professional life and raising children that you don't have time.

**Wes:** Tell us about your family.

**Seymour:** I have four children who I am very happy with. They're all doing reasonably well. My kids have grown up and gone through college, and are all, thank God, employed and reasonably happy in their work. Mindy lives nearby and is a biostatistician. Adam lives in Seattle and Dan and Georgia in Maryland, so I'm sort of in the middle and I get to see them on occasion. I'm a grandfather to Emma and Liam and to triplets that my youngest son fathered, Rachel, Eden, and Joshua. I've been married to my second wife for about 22 years, which has been a much happier marriage. My brother, Martin, who has been retired for a number of years, was a high school teacher in English and a counselor. He lives out on Long Island in New York.

**Wes:** Anything else you'd like to say?

**Seymour:** I have lived a reasonably happy life. I've had excellent students, some I'm sure will have better legacies to leave than I. I'm counting on that. I was one who tried to push a certain view that I had about what Statistics should do, in other words, prediction and observables and hoped that this would be helpful in statistical inference, in science, and in social science.

## ACKNOWLEDGEMENTS

Aelise Houx, retired from the Division of Statistics at the University of California, Davis, was kind enough to type the conversation from the taped interview. We greatly appreciate her efforts. We also thank Anne Geisser, Jessica Utts, Marvin Zelen and Martin Geisser, for their help in editing the conversation for accuracy. Unfortunately, Seymour's health did not permit him to participate in the editing. We also thank Ed Bedrick for his comments.